\RequirePackage{fix-cm}  
\documentclass[aps,prl,showpacs,amsmath,amssymb,amsfonts,superscriptaddress,lengthcheck,twocolumn]{revtex4-1}
\usepackage{graphicx}
\usepackage{subfigure}
\usepackage{amsthm}
\usepackage{verbatim}
\usepackage{dcolumn}
\usepackage{bm}
\usepackage{epsf}
\usepackage{color}
\usepackage[colorlinks=true,citecolor=blue,linkcolor=blue,urlcolor=blue]{hyperref}%
\newcommand{\bra}[1]{\left\langle #1\right|}
\newcommand{\ket}[1]{\left|#1\right\rangle}
\newcommand{\braket}[2]{\left\langle #1|#2\right\rangle}

\newcommand{\tr}[1]{\mathrm{tr}\left\{#1\right\}}

\newcommand{\la}{\left\langle}
\newcommand{\ra}{\right\rangle}

\newcommand{\de}[1]{\delta\left(#1\right)}

\newcommand{\e}[1]{\exp{\left(#1\right)}}
\newcommand{\lo}[1]{\ln{\left(#1\right)}}
\newcommand{\id}{\mathbb{I}}

\newcommand{\bla}{bla\\bla\\bla\\bla\\bla}

\newcommand{\mbb}[1]{\mathbb{#1}}

\newcommand{\mrm}[1]{\mathrm{#1}}

\begin{document}
\title{Quantum fluctuation theorem to benchmark quantum annealers}
\author{Bart\l{}omiej Gardas}
\email{bartek.gardas@gmail.com}
\affiliation{Theoretical Division, LANL, Los Alamos, New Mexico 87545, USA}             
\affiliation{Institute of Physics, University of Silesia, 40-007 Katowice, Poland}     
\affiliation{Instytut Fizyki Uniwersytetu Jagiello\'nskiego, ul. {\L}ojasiewicza 11, PL-30-348 Krak\'ow, Poland}  
\author{Sebastian Deffner}
\email{deffner@umbc.edu}
\affiliation{Department of Physics, University of Maryland Baltimore County, Baltimore, MD 21250, USA}

\date{\today}
\begin{abstract}	
Near term quantum hardware promises unprecedented computational advantage. Crucial in its development is the characterization and minimization of computational errors. We propose the use of the quantum fluctuation theorem to benchmark the performance of quantum annealers. This versatile tool provides simple means to determine whether the quantum dynamics are unital, unitary, and adiabatic, or whether the system is prone to thermal noise. Our proposal is experimentally tested on two generations of the D-Wave machine, which illustrates the sensitivity of the fluctuation theorem to the smallest aberrations from ideal annealing. 
\end{abstract}

\maketitle

\paragraph*{Introduction.} It is generally expected that for specific tasks already the first generations of quantum computers will have the potential to significantly outperform classical hardware  \cite{Boixo2016}. Loosely speaking this so-called quantum supremacy relies  on the fact that the quantum computational space is exponentially larger than the classical logical state space \cite{Nielsen2010}. 

In classical computers, Landauer's principle assigns  a characteristic thermodynamic cost to processed information -- namely to erase (or write) one bit of information at least $k_B T \lo{2}$ of thermodynamic work (or heat) have to be expended~\cite{landauer_1961,landauer_1991,Berut2012,Deffner2013,Boyd2016}. Recent years have seen the rapid advent of thermodynamics of information~\cite{Sagawa2009,Deffner2013,Horowitz2014,Parrondo2015,Strasberg2017}, which is a generalization of thermodynamics to small, information processing systems that typically operate far from equilibrium. In their description, tools and methods from stochastic thermodynamics have proven to be versatile and powerful. In particular, the fluctuations theorems enabled to generalize and specify Landauer's principle to a wide variety of systems~\cite{Sagawa2008,Horowitz2010,Sagawa2010}. 

In stochastic thermodynamics work is essentially a concept from classical mechanics, and it is given by a functional along a trajectory of the system \cite{Jarzynski1997,Jarzynski2011,Jarzynski2015}. For quantum systems the situation is significantly more involved, since quantum work is not an observable in the usual sense~\cite{Talkner2007,Campisi2011}. Thus, progress in the development of ``quantum thermodynamics of information'' has been hindered by the conceptual difficulties arising from identifying the appropriate definition of quantum work~\cite{Gardas15,Deffner2013EPL,Allahverdyan2014,Roncaglia2014,Hanggi2015,Talkner2016,Deffner2016}.

The most prominent approach relies on two projective measurements of the energy, one in the beginning and one at the end of the process~\cite{kurchan_2000,tasaki_2000}. If the systems is isolated, \emph{i.e.}, if the dynamics is at least unital, then the difference of the measurement outcomes can be considered as thermodynamic work performed during the process~\cite{Talkner2007,Campisi2011,Albash2013,Rastegin2013,Jarzynski2015PRX,Gardas2016}. This notion of quantum work fulfills a quantum version of the Jarzynski equality~\cite{kurchan_2000,tasaki_2000}, which has been verified in several experiments~\cite{Batalhao2014,An2015,Smith2018}. However, the question remains whether such a notion of quantum work, and the corresponding fluctuation theorem is useful in the sense that something can be ``learned'' about the system that one did not know already -- before the experiment was performed.

Since projective measurements are an important tool in quantum information and quantum computation~\cite{Nielsen2010}, it was only natural to generalize the quantum Jarzynski equality to a more general fluctuation theorem for arbitrary observables. The resulting theorem, $\la \e{-\Delta\omega}\ra=\gamma$, is formulated for the information production, $\Delta\omega$, during arbitrary quantum processes~\cite{Vedral2012,deffner2012,Manzano2015}. Here, $\gamma$ is the quantum efficacy that encodes the compatibility of the initial state, the observable, and the quantum map, and it is closely related to Holevo's bound~\cite{deffner2012}. Remarkably, $\gamma$ becomes a constant independent of the details of the process for unital quantum channels~\cite{deffner2012,Rastegin2013}. Physically, unital dynamics can be understood as systems which are subject to information loss due to decoherence, but do not experience thermal noise~\cite{Smith2018}.

In the following, we propose and exemplify the applicability of the general quantum fluctuation theorem in the characterization of the performance of quantum simulators. In particular, we show that the fluctuation theorem~\cite{deffner2012} can be utilized to test whether the quantum simulator is prone to noise induced computational errors. To this end, we will see that (i) if the quantum simulators is isolated from thermal noise, \emph{i.e.}, its dynamics is unital the fluctuation theorem is fulfilled, (ii) if the dynamics are unitary and adiabatic the probability density function of $\Delta\Omega$ is a $\delta$-function, \emph{i.e.}, a unique outcome of the computation is obtained.

Our conceptual proposal was successfully tested on two generations  of the D-Wave machine (2X and 2000Q). Our findings allow to quantify the resulting error rates from decoherence and other noise sources. Thus we show that the quantum fluctuation theorem and its related methods provide a powerful tool in the characterization of quantum computing hardware and their computational accuracy.

%
\paragraph*{General information fluctuation relation.}
\begin{figure}
	\includegraphics[width=\columnwidth]{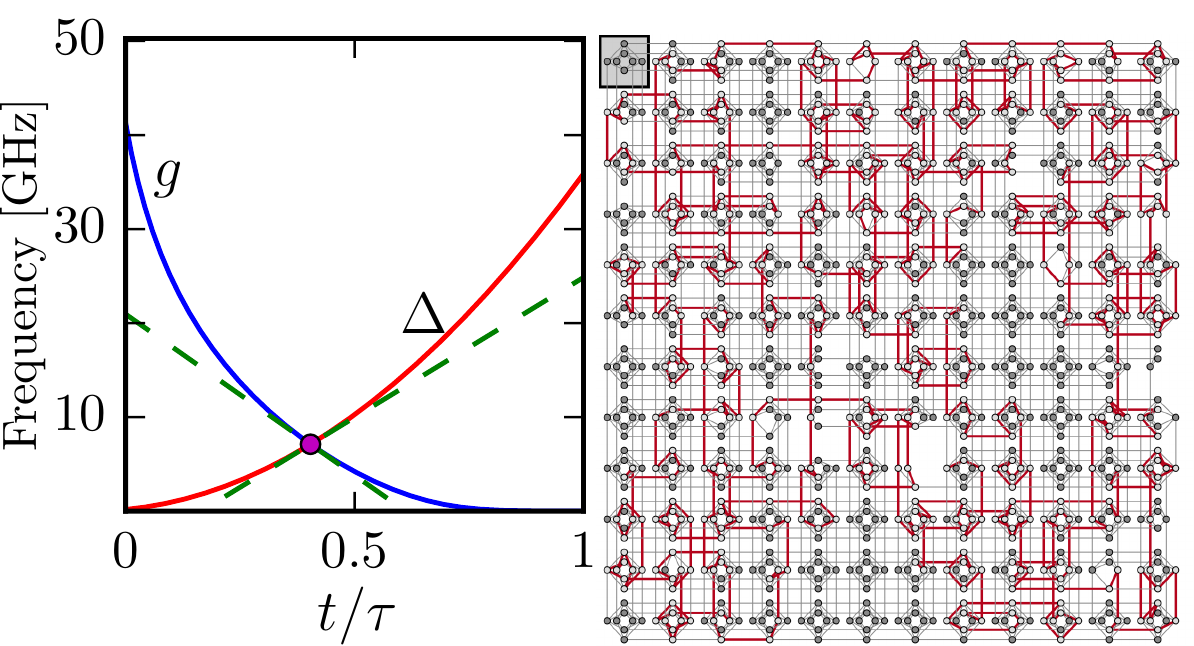}
	\caption{
		A typical annealing protocol for the quantum Ising chain implemented on the chimera graph (right panel). 
		The red lines are active couplings between qubits. The annealing time reads $\tau$.
	}
	\label{fig:chimera}
\end{figure}

To begin we briefly review notions of the general quantum fluctuation theorem~\cite{deffner2012} and establish notations. Information about the state of a quantum  system, $\rho_0$, can be obtained by performing measurements of observables. At $t=0$, \emph{i.e.}, to initiate the computation, we measure $\Omega^\mrm{i}=\sum_m \omega_m^\mrm{i}\Pi_m^\mrm{i}$. Note that the eigenvalues $\omega_m^\mrm{i}$ can be degenerate, and hence the projectors $\Pi_m^\mrm{i}$ may have rank greater than one. Typically $\rho_0$ and $\Omega^\mrm{i}$ do not commute, and thus $\rho_0$ suffers from a measurement back action. \cite{Nielsen2010}. Accounting for all possible measurement outcomes, the statistics after the measurement are given by the weighted average of all projections, 
\begin{equation}
\label{q01}
M^\mrm{i}[\rho_0] = \sum_m \Pi_m^\mrm{i}\, \rho_0\, \Pi_m^\mrm{i}\,.
\end{equation} 
After measuring $\omega_m^\mrm{i}$, the quantum systems undergoes a generic time evolution over time $\tau$ which we denote by $\mbb{E}_\tau$. At time $t=\tau$ a second measurement of observable $\Omega^\mrm{f} = \sum_n \omega_n^\mrm{f} \Pi_{n}^\mrm{f}$ is performed Accordingly, the transition probability $p_{m\rightarrow n}$ reads \cite{deffner2012}
\begin{equation}
\label{q02}
p_{m\rightarrow n}= \tr{\Pi_n^\mrm{f}\, \mbb{E}_\tau \left[ \Pi_m^\mrm{i} \rho_0\Pi_m^\mrm{i}\right]}\,.
\end{equation}
Our main object of interested is the probability distribution of all possible measurement outcomes, $\mathcal{P}\left( \Delta \omega\right)$, which we can write as~\cite{deffner2012}
\begin{equation}
\label{q03}
\mathcal{P}\left( \Delta \omega\right)=\sum\limits_{m,n} \de{\Delta \omega-\Delta \omega_{n,m}}\, p_{m\rightarrow n}\,,
\end{equation}
where $\omega_{n,m}\equiv\omega_n^\mrm{f} - \omega_m^\mrm{i}$. It is then easy to see \cite{deffner2012} 
\begin{equation}
\label{FR}
\la \e{- \Delta \omega} \ra = \gamma\,.
\end{equation}
The quantum efficacy $\gamma$ plays a crucial role in the following discussion and it can be written as
\begin{equation}
\label{q06}
\gamma = \tr{\e{-\Omega^\mrm{f}}\, \mbb{E}_\tau \left[M^\mrm{i}(\rho_0) \e{ \Omega^\mrm{i}} \right] }\,.
\end{equation}
Note that $\gamma$ is constant, (\emph{i.e.} process independent), for unital quantum dynamics~\cite{deffner2012}, in particular $\gamma$ becomes independent of the process length $\tau$. For such cases, it is always possible to redefine $\Omega^\mrm{i}$ and $\Omega^\mrm{f}$ such that $\gamma=1$. Thus, one could say that Eq.~\eqref{FR} constitutes a general fluctuation theorem for unital dynamics. On the contrary, for non-unital dynamics the right hand side depends on the details of the dynamics, and thus Eq.~\eqref{FR} is not fluctuation theorem in the strict sense of stochastic thermodynamics \cite{Seifert2008}.

\paragraph*{Fluctuation relation for the ideal quantum annealer.}

  \begin{figure}
  	\includegraphics[width=.5\textwidth]{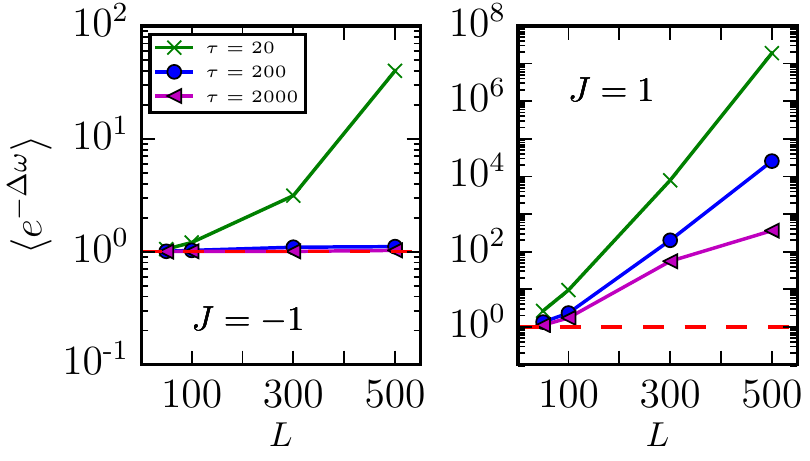}
  	\caption{\label{fig:FT1} Test of the quantum fluctuation~\eqref{FR} for the work distributions from Fig.~\ref{fig:results}a). Note that $\la \e{-\Delta\omega}\ra=1$ for all $\tau$ (depicted as red dotted line) signifies unital dynamics.} 
  	\label{fig:FT2}
  \end{figure}
  
We will now see that, on the one hand, the quantum fluctuation relation~(\ref{FR}) provides simple means to benchmark the performance of the hardware. On the other hand, quantum annealers such as the D-Wave machine provide optimal testing grounds to verify fluctuation relations in a quantum many body setup. 
\begin{figure*}[th!]
	\includegraphics[width=\textwidth]{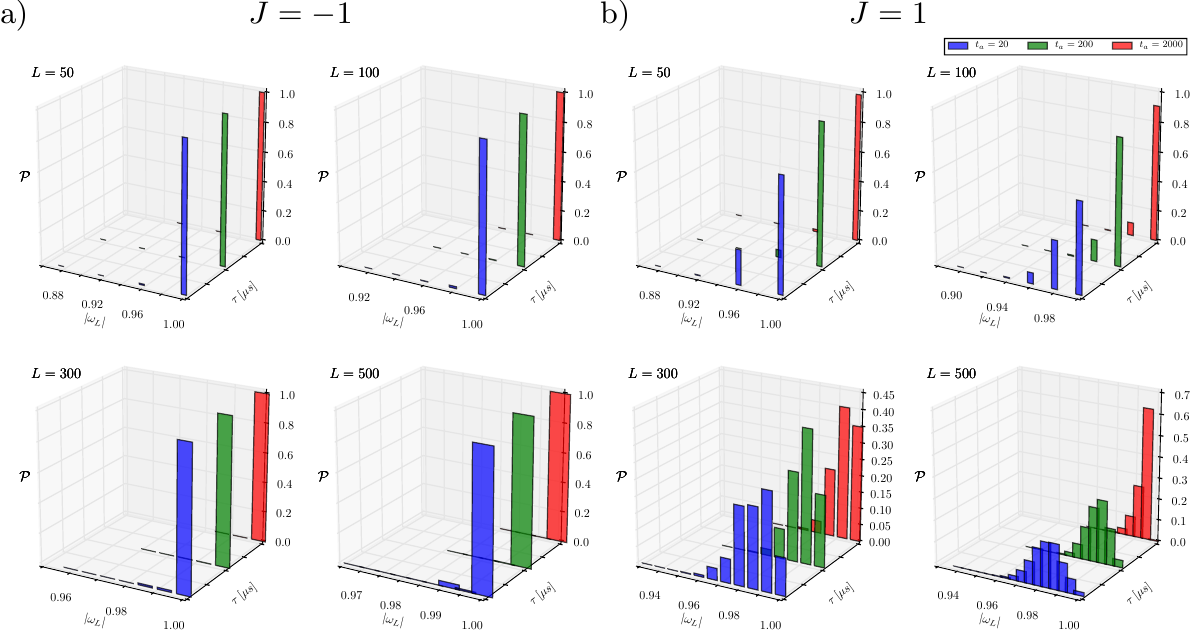}
	\caption{
		Distribution $\mathcal{P}(\Delta\omega)$ -- \eqref{q03} for the quantum Ising chain~(\ref{eq:ising}) implemented on a D-Wave $2$X annealer. Plot $a)$ and $b)$ shows the final results for $J=-1$ (antiferromagnetic) and $J=1$ (ferromagnetic) cases, respectively. To obtain each distribution $\mathcal{P}(\Delta\omega)$ the experiment was repeated $N=10^6$. Error bars are negligible and thus not shown in the plots. The renormalized energy is given by $\omega_L = \omega/(L-1)$, where $L$ is the length of a randomly chosen Ising chain. 
	}
	\label{fig:results}
\end{figure*}

To this end, we will assume for the remainder of the discussion  that the quantum system is described by the quantum Ising model in transverse field~\cite{Dorner2005},
\begin{equation} 
\label{eq:ising}
H(t)/(2\pi\hbar) = -g(t)\sum_{i=1}^{L}\sigma^x_i-\Delta(t)\sum_{i=1}^{L-1}J_i\sigma^z_i\sigma^z_{i+1}\,.
\end{equation}
Although, the current generation of quantum annealers can implement more general many body systems~\cite{Lanting14}, we focus on the simple one dimensional case for the sake of simplicity~\cite{Nishimori98}. An implementation of the latter Hamiltonian on the D-Wave machine is depicted in Fig.~\ref{fig:chimera}a. On this platform, users can choose couplings $J_i$ and longitudinal magnetic field $h_i$, which in our case are all zero. In general, however, one can \emph{not} control the annealing process by manipulating $g(t)$ and $\Delta(t)$. In the ideal quantum annealer the quantum Ising chain~\eqref{eq:ising} undergoes unitary and adiabatic dynamics, while $\Delta(t)$ 
is varied from $\Delta(0)\approx 0$ to $\Delta(\tau) \gg 0$, and $g(t)$ from $g(0) \gg 0 $ to $g(\tau)\approx 0$ (cf. Fig.~\ref{fig:chimera}a). 

The obvious choice for the observables is the (customary renormalized) Hamiltonian in the beginning and the end of the computation, $\Omega_i=H(0)/[2\pi\hbar g(0)]-\id$ and $\Omega_f=H(\tau)/[2\pi\hbar J\Delta(\tau)]$. Consequently, we have
	\begin{equation}
	\label{s1}
       \Omega_{\mrm{i}} = \sum_{i=n}^{L}\sigma^x_n - \id \quad \text{and} \quad 
       \Omega_{\mrm{f}} = \sum_{n=1}^{L-1}\sigma^z_n\sigma^z_{n+1}\,,
	\end{equation}
where we included $\id$ in the definition of $\Omega_{\mrm{i}}$ to guarantee $\gamma=1$ for unital dynamics.

For the ideal computation, the initial state, $\rho_0$, is chosen to be given by $\rho_0=\ket{\boldsymbol{\rightarrow}}\bra{\boldsymbol{\rightarrow}}$, where $\ket{\boldsymbol{\rightarrow}}:=\ket{\cdots \rightarrow\rightarrow\rightarrow\cdots}$ is a non-degenerate, paramagnetic state -- the ground state of $H(0)$ (and thus of $\Omega_{\mrm{i}}$), where all spins are aligned along the $x$-direction. As a result,  
\begin{equation}
    \label{s2}
    M_i[\rho_0]=\rho_0 \quad \text{and} \quad \omega_{\mrm{i}}=L-1,
    \end{equation}
 as $\Omega_{\mrm{i}}$ and $H(0)$ commute by construction~\footnote{Unfortunately, the D-Wave system does \emph{not} allow us to test the accuracy of the initial preparation that leads to Eq.~(\ref{s2}). However, applying a strong enough magnetic field, it should fairly easy to prepare a state where all spins are aligned in one direction.}.

Moreover, if the quantum annealer is ideal, then the dynamics is not only unitary, but also adiabatic. In this case, we can write $\mathbb{E}_{\tau}\left[\rho\right] = U_{\tau}\rho U_{\tau}^{\dagger}$, where  
\begin{equation}
       U_{\tau} = \mathcal{T}_{>}\exp\left(-\frac{i}{\hbar} \int_0^{\tau} H(s)\, ds\right) 
 \end{equation}  
and as a result $\mathbb{E}_{\tau}\left[\rho_0\right] = \ket{\boldsymbol{f}}\bra{\boldsymbol{f}}$,
where $\ket{\boldsymbol{f}}$ is the final state, a defect-free state where all spins are aligned along 
the $z$-direction, \emph{i.e.} $\ket{\boldsymbol{\uparrow}}$ or $\ket{\boldsymbol{\downarrow}}$. Therefore, $\omega_{\mrm{f}}=\omega_{\mrm{i}}$.
 
In general, however, due to decoherence~\cite{zurek03}, dissipation~\cite{adolfo17} or other (hardware) issues that may occur~\cite{Young13}, the evolution may be neither unitary nor adiabatic. Nevertheless, for the annealer to perform robust computation its evolution, $\mathbb{E}_{\tau}$, has to map $\ket{\boldsymbol{\rightarrow}}$ onto $\ket{\boldsymbol{f}}$. 
Therefore, the quantum efficacy~\eqref{q06} simply becomes
    \begin{equation}
     \label{sgamma}
     \gamma = e^{-\Delta\omega} \braket{\boldsymbol{f}}{\boldsymbol{f}} = 1,
     \end{equation}
that is, a process independent quantity.    

 Since the system starts from its ground state, $\ket{\boldsymbol{\rightarrow}}$, we can further write
      \begin{equation}
       p_{m\rightarrow n} = \delta_{0,m}\,p_{n|m} =  \delta_{0,m}\,p_{n|0}, 
      \end{equation}
      where $p_{n|0}$ is the probability of measuring $\omega_n^{\mrm{f}}$, conditioned on having first measured the ground state. Since we assume the latter event to be certain, $p_{n|0}\equiv p_n$ is just the probability of measuring the final outcome $\omega_n$ (we dropped the superscript). Therefore,
      \begin{equation}
        \langle e^{-\Delta\omega}\rangle = 
        e^{-\Delta\omega}\,p_0 + \sum_{n\not = 0} e^{-\Delta\omega_n}\,p_n.
      \end{equation}
      Comparing this equation with  Eq.~(\ref{sgamma}) we finally obtain a condition that is verifiable experimentally: 
      \begin{equation}
      \label{fdist}
       p_n = \mathcal{P}(|\omega_n|) = 
      \begin{cases}
	      1 & \text{if} \quad |\omega_n| = L-1,  \\
	      0 & \text{otherwise.} 
      \end{cases}
      \end{equation}
      
The probability density function $ \mathcal{P}(|\omega_n|) $  is characteristic for every process that transforms one ground state of the Ising Hamiltonian~\eqref{eq:ising} into another. It is important to note that the quantum fluctuation theorem~(\ref{FR}) is valid for arbitrary duration $\tau$ -- any slow and fast processes. Therefore, even if a particular hardware does \emph{not} anneal the initial state adiabatically, but only unitally (which is not easy to verify experimentally) Eq.~(\ref{fdist}) still holds -- given that the computation 
starts and finishes in a ground state, as outlined above.

As an immediate consequences, every $\tau$-dependence of $\mathcal{P}$ must come from dissipation or decoherence. This is a clear indication that the hardware interacts with its environment in a way that cannot be neglected.

\paragraph*{Experimental test on the D-Wave machine.}
\begin{figure*}[th!]
	\includegraphics[width=\textwidth]{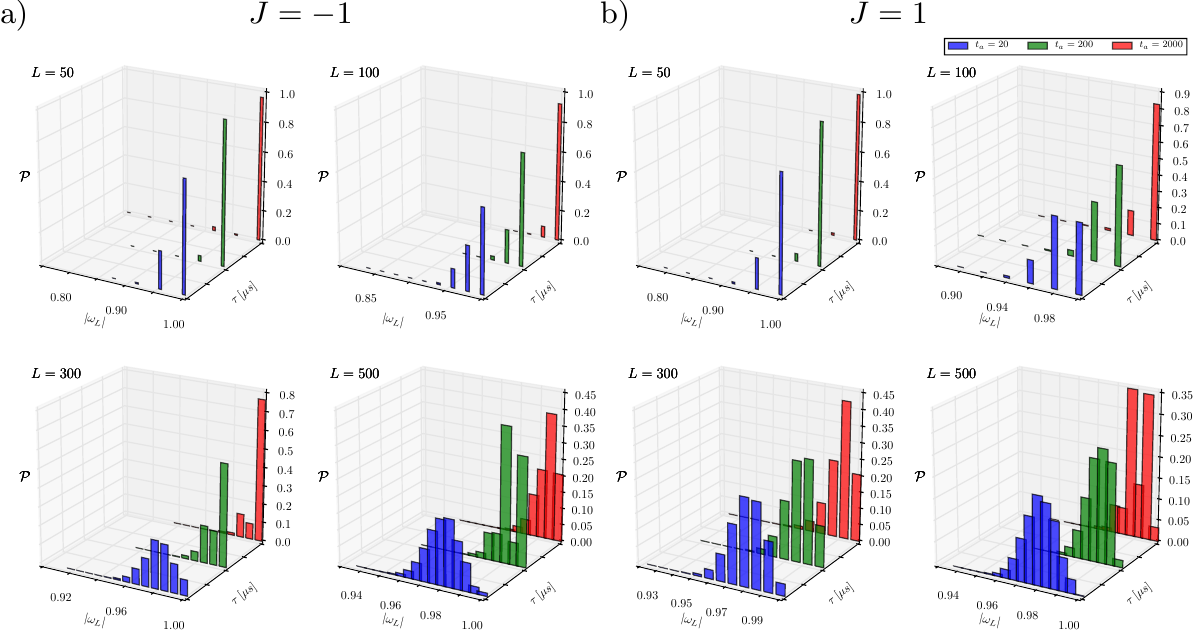}
	\caption{\label{fig:results2000Q}The same results as in Fig.~\ref{fig:results} but obtained from the newest generation of D-Wave quantum annealers ($2000$Q).
	}
\end{figure*}
We generated several work distributions $\mathcal{P}(|\omega_n|) $ -- \eqref{q03}  through ``annealing'' on two generations of the D-Wave machine ($2$X and $2000$Q), which implemented an Ising chain as encoded in Hamiltonian~(\ref{eq:ising}). All connections on the chimera graph have been chosen \emph{randomly}. A typical example is shown in Fig.~\ref{fig:chimera}b, where red lines indicate nonzero $zz$-interactions between qubits. The experiment was conducted $N=10^6$ times. Figure~\ref{fig:results} shows our final results obtained for different chain lengths $L$, couplings between qubits $J_i$ and annealing times $\tau$ on $2$X, and Fig.~\ref{fig:results2000Q} for $2000$Q . The current D-Wave solver reports the final state energy which is computed classically from the measured eigenstates of the individual qubits. In Fig.~\ref{fig:FT1} we show the resulting exponential averages, $\la \e{-\Delta\omega}\ra$.
   
\paragraph*{Discussion of the experimental findings.}
We observe, that there are cases for which the agreement is almost ideal. In particular, this is the case on $2$X for $J=-1$ and slow anneal times $\tau$, see Fig.~\ref{fig:results}. In this case the $ \mathcal{P}(|\omega_n|) $ is close to a Kronecker-delta, and the dynamics is unital, see Fig.~\ref{fig:FT1}. Note that the validity of the fluctuation theorem~\eqref{FR} is a very sensitive test to aberrations, since rare events and large fluctuations are exponentially weighted.

However, in the vast majority of cases  $\mathcal{P}(|\omega_n|) $ is far from our theoretical prediction~\eqref{fdist} and the dynamics is clearly not even unital, compare Fig.~\ref{fig:FT2}. Importantly, $\mathcal{P}$ clearly depends on $\tau$ indicating a large amount of computational errors are generated during the annealing. 

Interestingly, the D-Wave $2$X we tested~\footnote{This machine is based in Los Alamos National Laboratory} produces asymmetric results. The work distributions for ferromagnetic ($J>0$) and antiferromagnetic ($J<0$) couplings should be identical. On the other hand, the newest $2000$Q D-Wave machine exhibits less asymmetrical behavior, however, its overall performances is not as good as its predecessor's (see Fig~\ref{fig:results2000Q}). 

Complicated optimization problems involve both negative and positive values of the coupling matrix $J_{ij}$. That makes debugging ``asymmetric" quantum annealers a much harder task. Our proposal for benchmarking the hardware with the help of the quantum fluctuation theorem allows users to asses to what extent a particular hardware exhibits this unwanted behavior. Moreover, our test is capable of detecting any exponentially small departure from ``normal operation"  that may potentially result in a hard failure.  
We believe this to be the very first step to create \emph{fault tolerant} quantum hardware~\cite{Martinis15}.
     
As a final note, we emphasize that any departure from the  ideal distribution $\mathcal{P}$ (\ref{fdist}) for the Ising model indicates that the final state carries ``kinks'' (topological defects). Counting the exact number of such imperfections allows one to determine by how much the annealer misses the true ground state~\cite{defects17}. In a perfect quantum simulator this number should approach zero. 

The Ising model~(\ref{eq:ising}) undergoes a quantum phase transition~\cite{Dorner2005}. Near the critical point, \emph{i.e.}, at $t_c$ where $\Delta(t_c)=g(t_c)$, the gap  -- energy difference between the ground and a first accessible state -- scales like $1/L$. Thus, one could argue that the extra excitations come from a Kibble-Zurek like mechanism~\cite{Kibble76,Zurek85}. However, even the fastest 
quench ($\tau \sim 20 \,\mu s$) exceeds the adiabatic threshold~\cite{Dziarmaga2005},
\begin{equation}
\tau_{\mrm{ad}} \sim \frac{L^2}{\Delta(t_c)} \sim 10 \,\mu s\, , 
\end{equation}
for the system sizes of order $L\sim 10^2$.

\paragraph*{Concluding remarks.}
  
In the present analysis we have obtained several important results: (i) We have proposed a practical use and applicability of quantum fluctuation theorems. Namely, we have argued that the quantum fluctuation theorem can be used to benchmark the performance of quantum annealers. Our proposal was tested on two generations of the D-Wave machine. Thus, (ii) our results indicate the varying performance of distinct machines of the D-Wave hardware. Thus, our method can be used to identify underperforming machines, which are in need of re-calibration.  Finally, (iii) almost as a byproduct we  have performed the first experiments and verification of quantum fluctuation theorems in a many particle system.

\begin{acknowledgements}
We appreciate discussions with Edward Dahl of D-Wave Systems. Work B.G. was supported by Narodowe Centrum Nauki under Project No. 2016/20/S/ST2/00152. This research was supported in part by PL-Grid Infrastructure. SD acknowledges support by the U.S. National Science Foundation under Grant No. CHE-1648973.
\end{acknowledgements}

\bibliography{dwave}
\end{document}